\documentclass[reprint]{elsarticle}

\journal{NIM Section A}
\usepackage{amssymb}
\usepackage{amsmath}

\usepackage{graphicx}% Include figure files
\usepackage{dcolumn}% Align table columns on decimal point
\usepackage{bm}% bold math
\usepackage{natbib}
\newcommand{\vbb}{$0 \nu \beta \beta $}
\newcommand{\vvbb}{$2 \nu \beta \beta $}
\newcommand{\C}{$^{10}$C}
\newcommand{\Xe}{$^{136}$Xe}

\begin{document}

\begin{frontmatter}

\title{Suppression of Cosmic Muon Spallation Backgrounds in Liquid Scintillator Detectors Using Convolutional Neural Networks}

\author[BU]{A. Li\corref{cor1}}
\ead{liaobo77@bu.edu} 
\cortext[cor1]{Corresponding author}
\author[UChicago]{A. Elagin}
\author[MIT]{S. Fraker}
\author[BU]{C. Grant}
\author[MIT]{L. Winslow}
\address[BU]{Boston University, Department of Physics, 590 Commonwealth Avenue, Boston, MA 02215, USA}
\address[UChicago]{Enrico Fermi Institute, University of Chicago, 5640 S. Ellis Ave, Chicago, IL, 60637}
\address[MIT]{Laboratory for Nuclear science, Massachusetts Institute of Technology, 77 Massachusetts Avenue, Cambridge, MA 02139, USA}

\begin{abstract}
Cosmic muon spallation backgrounds are ubiquitous in low-background experiments. For liquid scintillator-based experiments searching for neutrinoless double-beta decay, the spallation product \C\ is an important background in the region of interest between 2-3\,MeV and determines the depth requirement for the experiment. We have developed an algorithm based on a convolutional neural network that uses the temporal and spatial correlations in light emissions to identify \C~background events. With a typical kiloton-scale detector configuration like the KamLAND detector, we find that the algorithm is capable of identifying 61.6\% of the \C~ at 90\% signal acceptance. A detector with perfect light collection could identify 98.2\% at 90\% signal acceptance. The algorithm is independent of vertex and energy reconstruction, so it is complementary to current methods and can be expanded to other background sources.
\end{abstract}

\begin{keyword}
%% keywords here, in the form: keyword \sep keyword
 liquid scintillator \sep neutrino detector \sep deep learning \sep spallation \sep neural network \sep background rejection
%% MSC codes here, in the form: \MSC code \sep code
%% or \MSC[2008] code \sep code (2000 is the default)

\end{keyword}

\end{frontmatter}

%%
%% Start line numbering here if you want
%%
% \linenumbers
%\maketitle

%% main text
\section{\label{sec:intro}Introduction}
Neutrinoless double-beta decay (\vbb) is a hypothetical decay process by which a nucleus ejects two electrons and no neutrinos, therefore violating lepton number by two units. The observation of this process would demonstrate that neutrinos are Majorana fermions, yield valuable insight into the mechanism behind neutrino mass generation, and support a theoretical framework for matter-antimatter asymmetry in the early universe. Experiments currently being planned will instrument up to a few tons of isotope aiming for \vbb\;half-life sensitivities between $10^{27} - 10^{28}$\,years~\cite{nEXO_PRC2018, LEGEND_AIPConf2017, giuliani_andrea_2018_1286915}.

Large liquid scintillator detectors are attractive for these searches because they offer cost effective scaling to large volumes and effective background reduction through self-shielding, spatial and temporal coincidence analyses, and pulse shape discrimination. The KamLAND-Zen experiment has shown the effectiveness of this technique by setting the most stringent limit on the \vbb\ half-life independent of isotope $T^{0\nu}_{1/2}>1.07\times10^{26}$ for \Xe~\cite{KLZ2016}. 

Due to the size of these detectors, the typical 20\% photocathode coverage, and $\gtrsim$1\,ns timing resolution, all energy depositions are assumed to originate from a single point in space and in time. However, we know that different particle species have characteristic topologies which change in time. Even with current detector performance, these could be used for particle identification and background reduction in a variety of analyses. Recent advancements in photodetector technology, with the advent of Large Area Picosecond Photodetectors (LAPPDs)~\cite{Timing_paper, Incom_paper}, decrease the photon arrival time uncertainty to $\leq 100$\,ps. This allows for particle identification by topology~\cite{harmonics2017} and may also permit the reconstruction of particle direction through the separation directional Cherenkov light from the abundant isotropic scintillation light~\cite{direction2014}. This has been demonstrated with muons~\cite{chessEPJC,chessPRC,bnl2016} and recently with $^{90}$Sr $\beta$-decays~\cite{flatdot2018}.

One of the primary cost drivers of \vbb\;experiments is the depth at which they must be located in order to minimize backgrounds due to cosmic muon spallation. In liquid scintillator detectors doped with \vbb\;isotopes with endpoints below 3\,MeV, such as \Xe $\,$($Q=2.458$ MeV), the critical long-lived light isotope is \C $\,$($Q=3.648$ MeV, $\tau_{1/2}=19.29$ s). In the current KamLAND-Zen result, the largest background is the two-neutrino double-beta decay (\vvbb) of \Xe, which is only reducible with improved energy resolution. In the absence of scintillator or photodetector upgrades, the largest reducible background is \C~\cite{KLZ2016}.  Looking at future multi-ton scintillator experiments, the increased size and shallower depth make \C\;the largest background for a \vbb\;search with the JUNO experiment~\cite{junoBB2015,juno0BB2017}.  The \C\;background will also inform the choice of the depth and location of the proposed THEIA experiment~\cite{theiaOld}.

The production of neutrons and light isotopes in muon spallation is an active area of study \cite{kamspall, borexinoSpall, doublechoozSpall, dayabayNeutron}. \C\;is a relatively common spallation product, predominantly made through $(\pi^+, np)$~\cite{kamspall}.  Results from the Borexino experiment suggest a three-fold-coincidence of muon, neutron, and $^{11}$C decay can be used to tag the $^{11}$C decay relative to the neutron capture vertex and muon track~\cite{galbiati2004, bellini2014}. This three-fold-coincidence, accompanied by a neutron in the final state, is also used to tag \C\; in the KamLAND-Zen analysis~\cite{KLZ2016}. However, muon spallation is inherently chaotic and the copious number of neutrons produced through this process lead to frequent periods of detector instability. This is especially true for the highest energy events, like showering muons, which produce most of the light isotopes. The problem can be addressed with improved electronics, but due to the large dynamic range between muon events (volt-level signals) and neutron events (millivolt-level signals), difficulties remain. The problem is further amplified by theoretical uncertainty in the fraction of \C~ production with neutron final states, and the spatial distribution relative to the muon track. For these reasons, an independent method of identifying \C~is useful for both reducing the background and verifying our understanding of the spallation process.

In this work we demonstrate that an algorithm based on a ten-layer convolutional neural network (CNN) developed for machine vision applications can effectively separate \C\;from \vbb\;events in a kilo-ton scale liquid scintillator detector, like the current KamLAND detector, without relying on muon or neutron coincidences. Since the technique is spatially invariant, it goes a step farther and is also independent of vertex reconstruction. We then perform a series of studies to understand what information the CNN is using in its discrimination and how the performance of the algorithm changes as we improve the KamLAND-like detector's performance. 

The paper is organized as follows. Section~\ref{sec:0vbb_vs_C10} describes the topological differences between \vbb-decay and \C~events that allow them to be distinguished by the algorithm. Section~\ref{sec:Detector_MC} provides the details of the detector Monte Carlo (MC) simulation and Section~\ref{sec:CNN} provides the details of the algorithm. These are followed by the results and conclusions in Sections~\ref{sec:Results} and~\ref{sec:Conclusion}.

\section{\label{sec:0vbb_vs_C10}Topology of \vbb-decay and \C~Events}
In a liquid scintillator detector, charged particles deposit energy which excites organic molecules. These molecules subsequently de-excite by releasing photons which are detected by single photon detectors like photomultiplier tubes (PMTs). The intrinsic timing of these processes is on the order of ns. Neutrino detectors with large volumes (diameters $>$10\,m) are typically instrumented with large PMTs with the capability to resolve photon arrival times on the order of 1-5~ns. 

Given the similarity between the time scales of the scintillation process and PMT readout, energy deposits are assumed to be point-like when reconstructing energies and vertices of physics events. This neglects two effects. Gamma-rays at these energies scatter multiple times with a mean free path on the order 10\,cm which leads to a smearing of the vertex in time and space. By comparison, electrons travel $\sim$1\,mm at these energies, but they are above Cherenkov threshold.  Therefore, the electrons produce some directional light which is not absorbed by the scintillation process. This information is encoded in the pattern of photons arriving at the PMTs and can be used to identify different categories of events.

In a large fraction of \vbb~events, the electrons exit the nucleus roughly back-to-back evenly dividing the available energy~\cite{Jenni}. For \Xe, this leads to two electrons each with a kinetic energy of roughly 1.23\,MeV. The Cherenkov threshold for electrons in this liquid scintillator is 0.16\,MeV. Examining the path of these electrons in Monte Carlo, we find that they travel $7.1\pm0.9$\, mm with a total distance from the origin of $5.6\pm1.0$\,mm in $26\pm$4 ps and drop below Cherenkov threshold after $24\pm3$\,ps. We also find that the final direction of the electron before it stops does not match the initial direction, however, the total scattering angle is small while Cherenkov light is emitted. 

\begin{figure}[htp]
\centering
\includegraphics [trim={1.1cm 0 0 0},clip, width=\columnwidth] {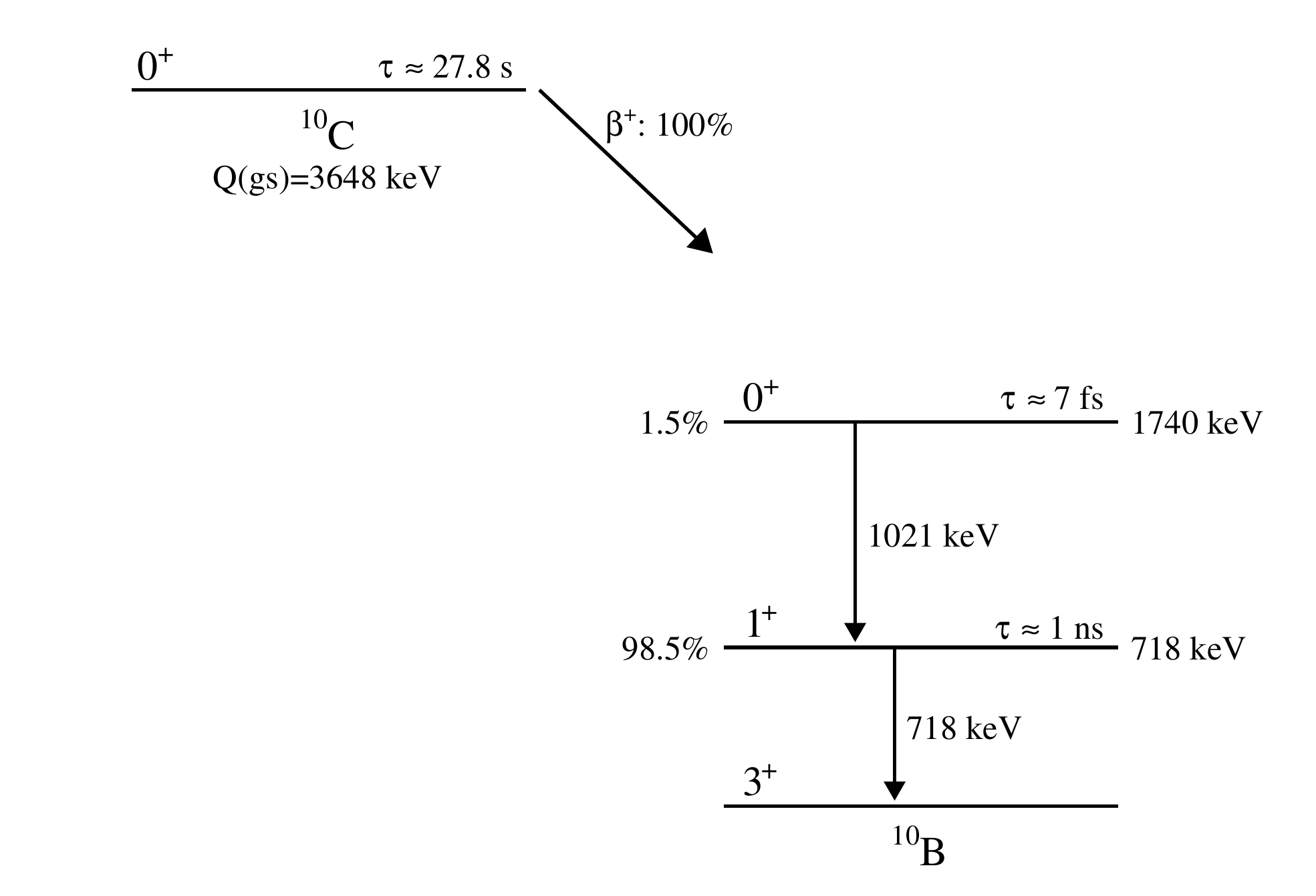}
\caption{Decay scheme of \C~\cite{JOuellet_C10_picture}. The final state of \C~events consists  of a positron and either one gamma with energy of 718~keV (98.5\%) or two gammas with energies of 718~keV and 1021~keV (1.5\%). }
\label{fig:C10}
\end{figure}

The decay of \C~is more complicated. \C~ is a $\beta^+$ decay that proceeds through one of two excited states as shown in Fig.~\ref{fig:C10}. The event observed by the detector is a combination of a positron, the two subsequent 511~keV annihilation gammas, and either one gamma with energy of 718~keV (98.5\%) or two gammas with energies of 718~keV and 1021~keV (1.5\%). In addition, the 718~keV excited state is long-lived with a lifetime of 1~ns. This is significant on the time scale of events in liquid scintillator detectors. 

The defining feature of the \vbb~is the two electrons above Cherenkov threshold and the ring-like features that algorithms, either traditional algorithms based on spherical harmonics or those based on machine learning, should pick out. Since the number of Cherenkov photons is low, the positron from \C~could mimic some of these features. However, the \C~vertex is significantly smeared in time and space by the three or more gammas in the final state. In the region of interest for \Xe, the positron has an energy of $\sim$0.7~MeV. Examining the path of the positron in Monte Carlo, we find that they travel $3.6\pm0.6$\,mm for a total offset from the origin of $2.5\pm0.6$ mm in $13\pm2$\,ps. It takes $11\pm2$\,ps for the positron to fall below Cherenkov threshold.

\section{\label{sec:Detector_MC} Detector Simulation}

These studies are performed using a Geant4-based Monte Carlo (MC) with a simplified 6.5\,m radius sphere of liquid scintillator. The properties of the liquid scintillator are chosen to match those of KamLAND. We record all photons that reach the outer surface and apply post processing to account for a variety of light collection configurations. This simulation is the same as those used in Ref.~\cite{harmonics2017} and Ref.~\cite{direction2014} where more details can be found on the modeling of the LS. 

Two types of events are generated for this study: \vbb~decay of $^{136}$Xe and the $\beta^{+}$~decay of $^{10}$C background. The kinematics of \vbb~decay events are simulated using a custom MC event generator with momentum and angle-dependent phase space factors from~\cite{Jenni}. \C~events are simulated using the default isotope decay generator in GEANT4~\cite{geant4one,geant4two}. This correctly accounts for the long-lived first excited state of $^{10}$B, but does not include the formation of positronium.

\begin{figure}[htp]
\centering
\includegraphics [width=8.8cm] {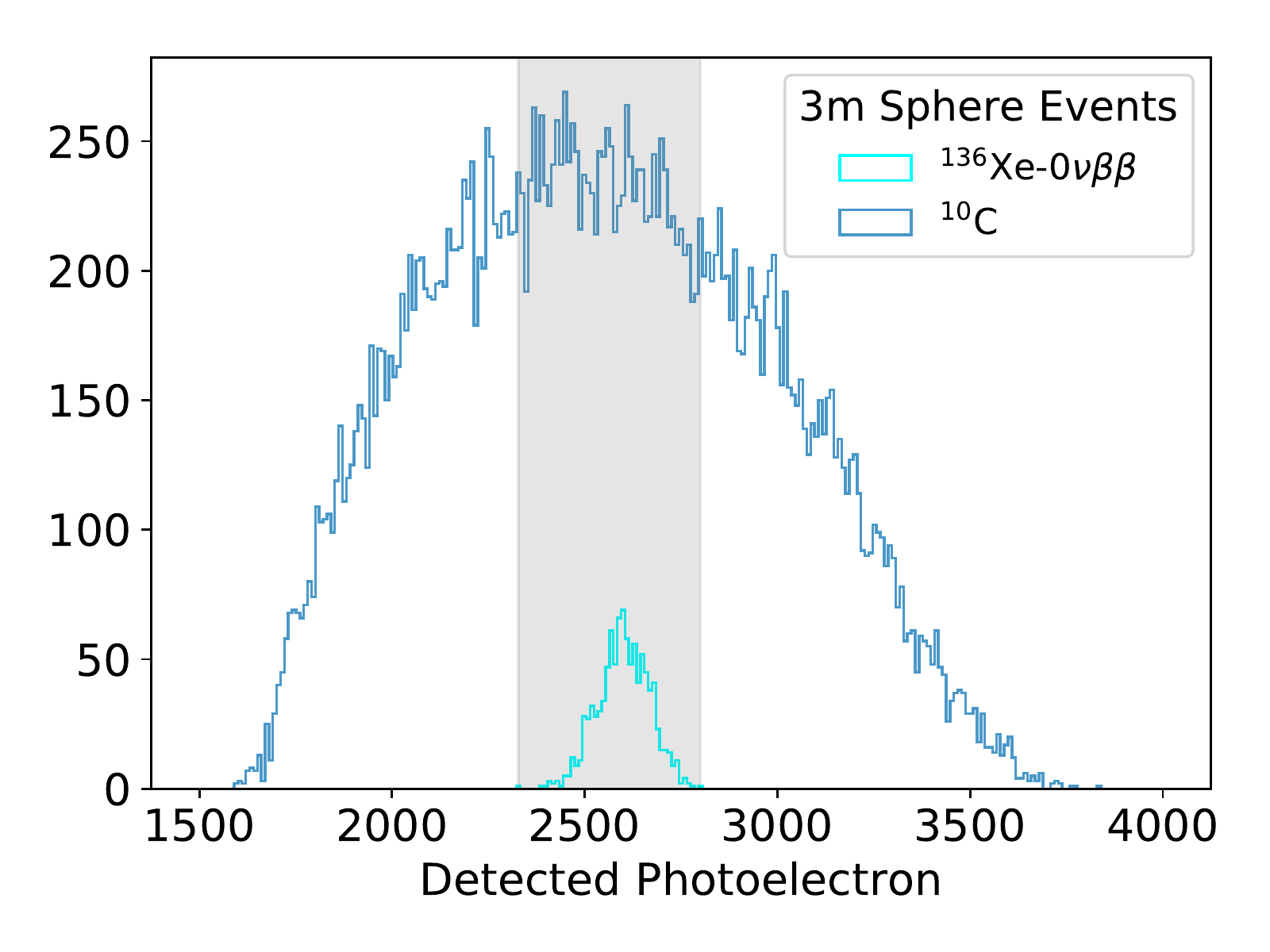}
\caption{The detected photoelectrons, assuming 100\% photocathode coverage and QE, for $^{136}$Xe \vbb\, decays and $^{10}$C $\beta^+$ decays generated inside a sphere with 3 m radius.  The gray band indicates the energy region of interest for \vbb.}
\label{fig:edep}
\end{figure}

The energy spectrum of both event types is shown in Fig.~\ref{fig:edep} in terms of detected photo-electrons for a detector with perfect light collection. Since  $^{10}$C event has a much broader spectrum comparing to $^{136}$Xe signal, an energy cut is placed at 2.2\,MeV to 2.7\,MeV prior to training in order to remove background events outside region of interest.

\begin{figure}[t!]
\centering
\includegraphics [trim={0.1cm 0 1cm 1cm},clip,width=8.8cm] {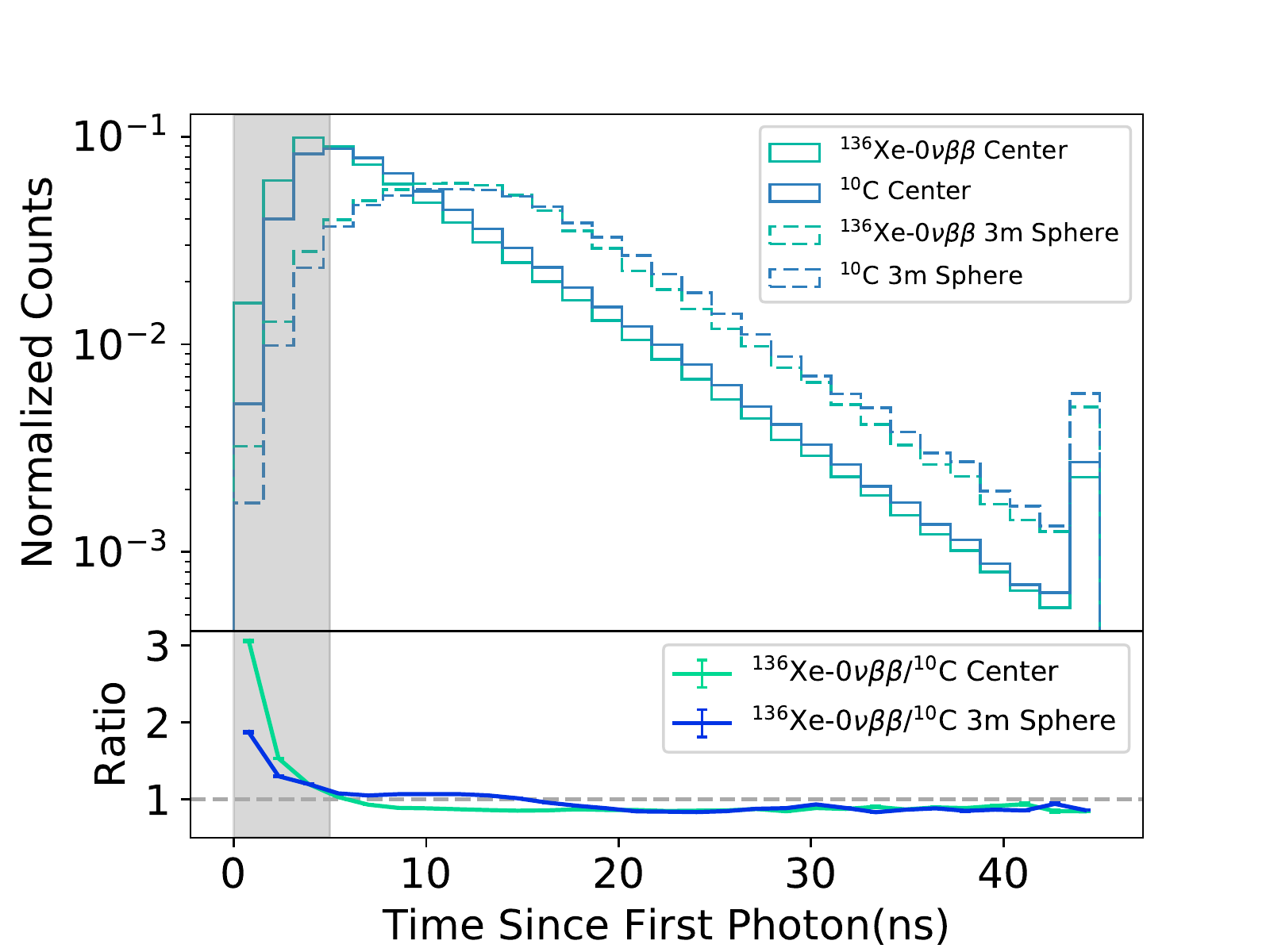}
\caption{TOP: Timing Profile of incoming photons for signal and background events. BOTTOM: Signal/Background ratio of timing profile histograms. Each bin represents a 1.5ns detector snapshot, taken from the detector's sampling rate. The grey band indicates the period of major discrepancy, thus the characteristics of signal/background learned by the network.  The dashed line represents a signal/background ratio equal to unity.}
\label{fig:timing}
\end{figure}

The simulation of these events are distributed either at the exact center of the detector, or uniformly within a 3-meter diameter spherical volume located at the center of the detector\footnote{Unless otherwise specified, the two types of event distributions will be referred as \textit{center} and \textit{3m Sphere} in this text.}. The latter matches the dimensions of the KamLAND-Zen mini-balloon, which contains the $^{136}$Xe-doped scintillator. Fig.~\ref{fig:timing} compares PE arrival times between \vbb-decay and \C~events. The time smearing has been included but the PMT quantum efficiency is added in later steps. In the events at the center, the excess early light from \vbb~and late light from \C~is evident. This pattern is repeated in the 3m sphere events but the shift in arrival time due to the vertex positions washes out some of the features. We note that the inclusion of positronium in the model would further increase the late light contribution in \C~decays, improving the results of this study.

\subsection{\label{sec:clock}Clock Latching Algorithm}
\begin{figure}[tt!]
\centering
\includegraphics [trim={3.5cm 1cm 1cm 0cm},clip, width=1.0\columnwidth] {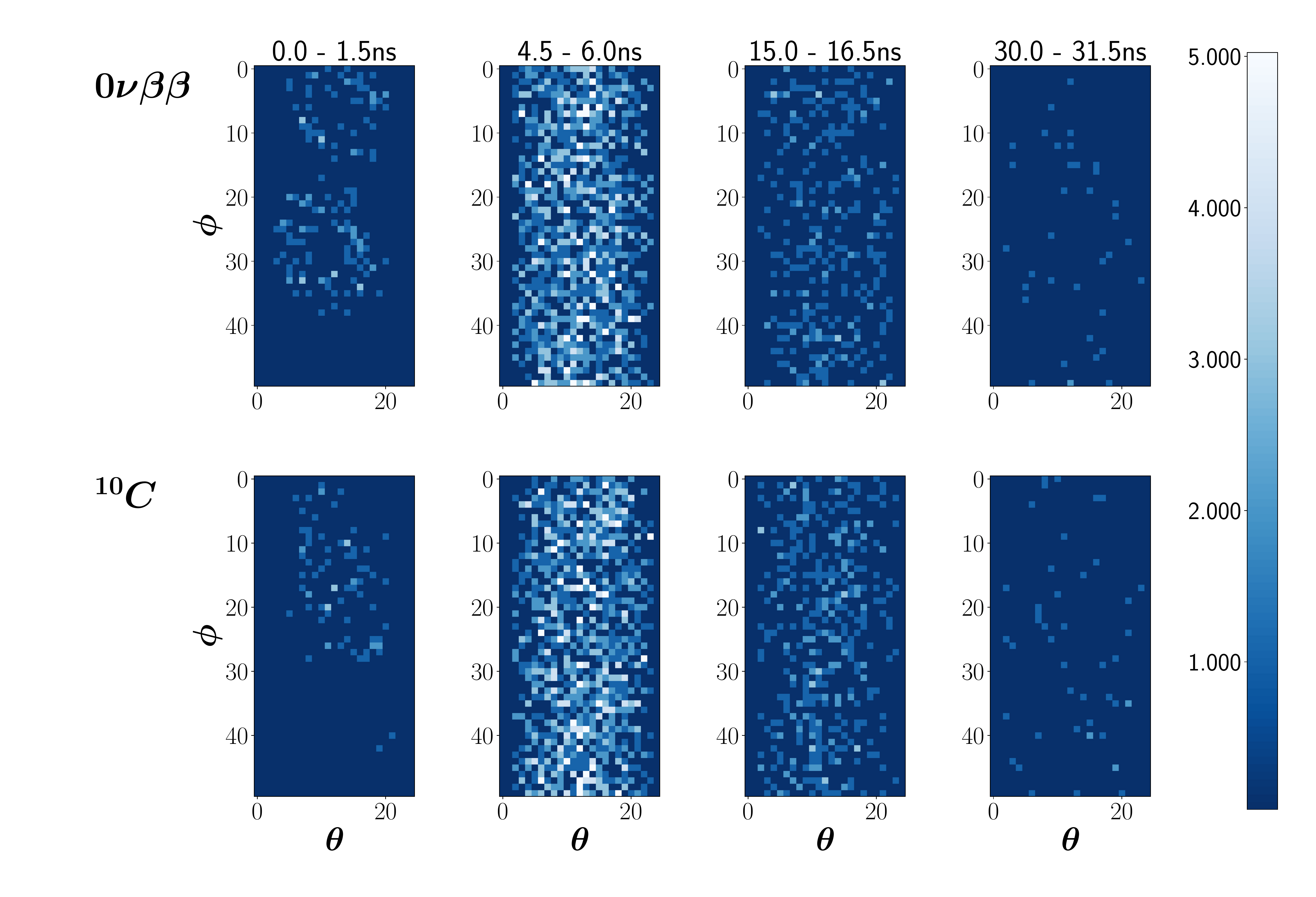}
\caption{Time evolution of the PMT hit-map over four example time bins for a \vbb\, event and a $^{10}$C event at the center of the detector.}
\label{fig:timeEvol}
\end{figure}

The photon arrival time is an important parameter of the simulation, so we model the digitization of the signal using a clock latching algorithm. Before latching, the hit time of each photon is smeared by a Gaussian Probability Density Function with a 1\,ns standard deviation in order to model the performance of these large PMTs. The sample time of the KamLAND electronics is 1.5\,ns. When the first unsmeared photon reaches the 6.5\,m boundary of the simulation, a clock will start ticking at an 1.5\,ns interval. Any photon that arrives after the current clock tick but prior to next tick will be latched to this tick. A total of 30 ticks, 45\,ns after the first photon, will be used to record the input event. Furthermore, 4 additional channels are added prior to zero time in order to take into account the backward time smearing of photon. A 2D image of the detector is then formed for each tick. We call each of these images a channel. Each event contains 34 channels, 33 from -6 to 45\,ns in 1.5\,ns increments and one larger bin for the remaining late photons. Fig.~\ref{fig:timeEvol} contains four example channels for \vbb~and \C. The distinctive double Cherenkov rings from \vbb~decay are visible in the first channel. The higher light levels from \C~are seen in the example channels corresponding to later times. 

\subsection{\label{sec:greydisk}Gray Disk PMT Model}

The photo-coverage of the detector is introduced by circumscribing a circular region around each PMT location, known as a \textit{gray disk}. We used the KamLAND PMT locations for the 1,325 17-inch PMTs as a representative PMT layout\cite{kamPMT}. The total gray disk area was adjusted to yield the desired photo-coverage. When a photon is produced in the MC, it is associated to the closest PMT. If this photon passes through the gray disk region of the associated PMT, one photon hit is recorded for that PMT position; otherwise, the photon is rejected. This method limits the total photocathode coverage to 40\%, since at that point the PMTs start overlapping.

\subsection{\label{sec:QE}Quantum Efficiency}
The PMT quantum efficiency (QE) is wavelength dependent and is another important simulation input. The QE dependence is accounted for during the event generation. In GEANT4, a QE bit associated with each photon is introduced to indicate whether or not it was recorded by the PMT. However, in this study, we need a varying QE to act as a pressure\footnote{In this context, the term pressure refers to the level of difficulty for a neural network to classify events. See also Sec.~\ref{sec:training}} parameter in order to demonstrate the neural network's performance. A stepwise QE cut is introduced to accommodate this requirement. If the QE bit indicates that a photon was detected, then the photon always passes the QE cut. Otherwise, photons are randomly rejected based on the desired QE pressure. In the ideal situation the QE is 100\% and all photons will be indiscriminately recorded.

 For our baseline KamLAND-like detector model, the current model, we assume a QE of 23\% and photocathode coverage of 19.6\%. We define an example upgrade scenario where these parameters are roughly doubled to a QE of 36.2\% and photocathode coverage of 42\%. This is close to what is being proposed for the JUNO and upgraded KamLAND experiments.  
 
\section{\label{sec:CNN}Event Classification Algorithm}
CNNs are a type of deep neural networks commonly used in the field of computer vision. To perform classification tasks using CNNs, a pixelized image is processed through several so-called {\it layers}, each containing a linear transformation step, followed by a non-linear activation. Among these layers, convolutional layers play the most important role. As the name implies they involve  the application of a convolution operation between two functions. For continuous functions, it represents the Fourier transform of the products. Applying this procedure to a 2D discrete surface gives rise to the convolutional layer. 

The convolutional filter, the kernel for the convolutional layer, is a fix-sized grid with specific values assigned to each block. The filter is scanned throughout the image body, and each image pixel is multiplied by a filter weight.  Finally, the convolution operation is completed by summing the element-wise multiplications.

The convolutional layer is also capable of taking information from multiple channels. In this case, the convolution operation is conducted separately over each channel, and the output values are summed and fed into the next layer.  For example, when a CNN is used to classify photographs, the input contains four channels:  Red(R), Green(G), Blue(B), and Grey Scale.  For this work, we have 34 time-based channels as described in Section~\ref{sec:clock}.

The output of the convolutional layer contains features that are fed into the fully connected layer. This layer is where the high level decisions are made, ending with an image being classified into one of several categories with an assigned probability. Other layers in the network structure include pooling layers and dropout layers. A pooling layer reduces the image dimensionality by extracting only the maximum value from each pooling filter to represent the image. It significantly increases the processing speed with almost no sacrifice of classification accuracy. The dropout layer prevents overfitting by randomly disabling neurons in the hidden layer with a predefined dropout rate~\cite{JMLR:v15:srivastava14a}. For extensive details on the functionality of CNNs, we refer the reader to Ref.~\cite{lecun_98}.

\subsection{\label{sec:network}Network Design}
\begin{figure}[bbb!]
\centering
\vspace{2mm}
\includegraphics [trim={1.5cm 1.5cm 0cm 1.5cm},clip, width=0.9\columnwidth] {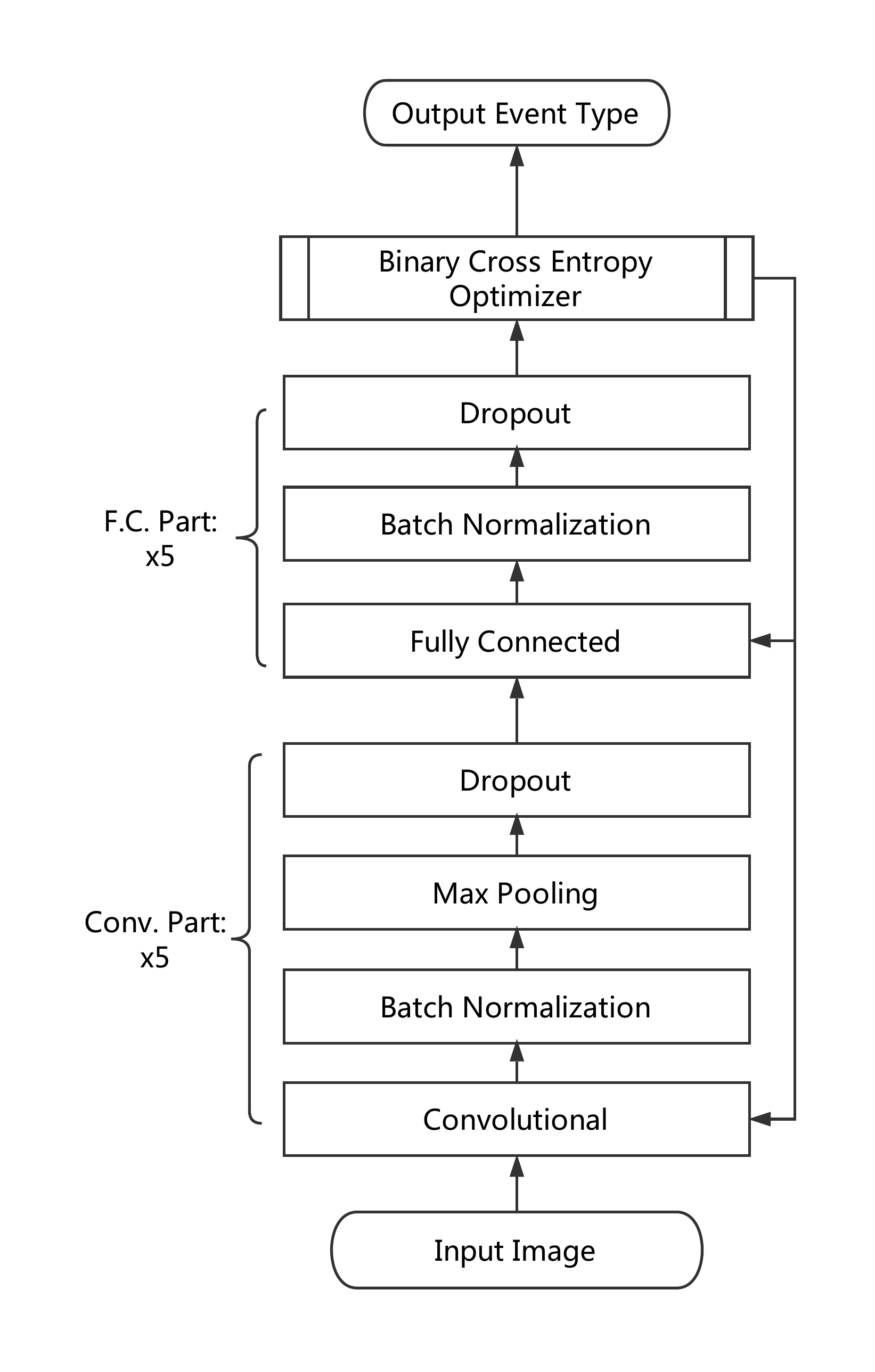}
\caption{Flow Diagram of the CNN.The Network is composed of Convolutional Part(Bottom) and Fully Connected Part(Top).} 
\label{fig:network}
\end{figure}

The CNN used in this work is implemented in Keras~\cite{chollet2015keras} with a Tensorflow backend~\cite{tensorflow2015-whitepaper}. The general outline of the network is shown in Fig.~\ref{fig:network}. It can be divided into a convolutional part and a fully connected part.

The convolutional part contains 5 sets of layers where each set includes a convolutional layer, a batch normalization layer, a pooling layer, and a dropout layer. For simplicity, one such set of layers is often simply referred to as a convolutional layer. Each time a pooling layer is introduced, the image reduces to 25\% of it's original size. Therefore, the convolution part is confined to five iterations.

The fully connected part also contains 5 sets of layers. Each set includes a fully connected layer, a batch normalization layer and a dropout layer. Again, this set of layers is often collectively referred to as a fully connected layer. With respect to the overall depth, the fully connected part is not limited by the image size or pooling layers. However, going too deep with the fully connected part will complicate the model and lead to overfitting. Dropout layers are inserted throughout the network in order to prevent this effect while decreasing the processing time~\cite{JMLR:v15:srivastava14a}.

Normalization of data plays an important role in classification tasks. Without normalization, the neural network will reveal some anomalous behavior, including non-convergance and high probability of misclassification. During a pre-training stage, each pixel is scaled to a value below unity. During the network design stage, several different normalization schemes were considered, including vector normalization, channel-wise standard scalar normalization, and batch normalization. For this work, batch normalization was chosen. This means that the normalization is performed for each incoming batch, transforming the input data such that the mean is zero with a standard deviation of one.

Hyperparameters refer to parameters in the network that are predefined before training, and stay constant throughout the training stage. Hyperparameters of neural networks define the structure of the network, and changing these parameters turns the neural network into a different model. Typical hyperparameters include, but are not limited to, the number of layers in the network, the number of nodes in each layers, the size of the filters, and the dropout rate.

A hyperparameter search can result in a significant improvement of performance~\cite{randomsearch_12}. For our model, we performed the search with hyperopt~\cite{Bergstra:2013:MSM:3042817.3042832}. Three hyperparameters are tuned to achieve the best performance, including number of nodes, number of fully connected layers, and dropout rate. We selected a preliminary training data set of 20,000 3m sphere events with the current detector configuration. We then search across a continuous range for the dropout rate between 0 and 1 and several discrete values for the number of fully connected layers and nodes. A random search~\cite{randomsearch_12} of 50 attempts is executed to determine the best result. The 50 trials are evaluated and compared for the best background rejection capability. After tuning, the validation accuracy increases from 49.7\% to 77.3\%.

\subsection{\label{sec:training}Training}
The algorithm is trained and validated on MC data sets, generated according to Section~\ref{sec:Detector_MC}. We study two different data sets: 70,000 centered events, and 50,000 3m sphere events. All events are stored in an 6 dimensional numpy array. The six dimensions are correspondingly: photocoverage pressure, QE pressure, event index, time channel, polar angle and azimuthal angle. The algorithm is trained and validated independently for each data set and each category contains equal amounts of \vbb~signal and \C\ background events. 

The data sets are separated into training and validation subsets with a 3:1 ratio. The network is trained on batches of 10 events over 30 training cycles. An RMSProp optimizer~\cite{tensorflow2015-whitepaper} is used to apply backward propagation optimization based on binary cross entropy~\cite{Mannor:2005:CEM:1102351.1102422}. Due to the large data volume, sparse matrix and batch generator technology is applied to reduce memory consumption. During training stage, a learning rate decay scheduler is incorporated to reduce systematic fluctuation. After training, the CNN is applied to the validation data set to study the out-of-sample performance. A bad validation rate in the presence of good training accuracy is indicative of overfitting and we do not see this effect.

While training the CNN on each event category, the photo-coverage and PMT QE were scanned over a wide range of values to better understand the CNN performance under different levels of classification difficulty.  These levels of difficulty are referred to as \textit{pressure}.  The photo-coverage was allowed to vary from 20\% to 40\% and QE from 23\% to 56\%. A total of 99 CNN models were trained and evaluated for the pressure maps.

\section{\label{sec:Results}Results}
\begin{figure}[hh!]
\centering
\includegraphics [width=8cm] {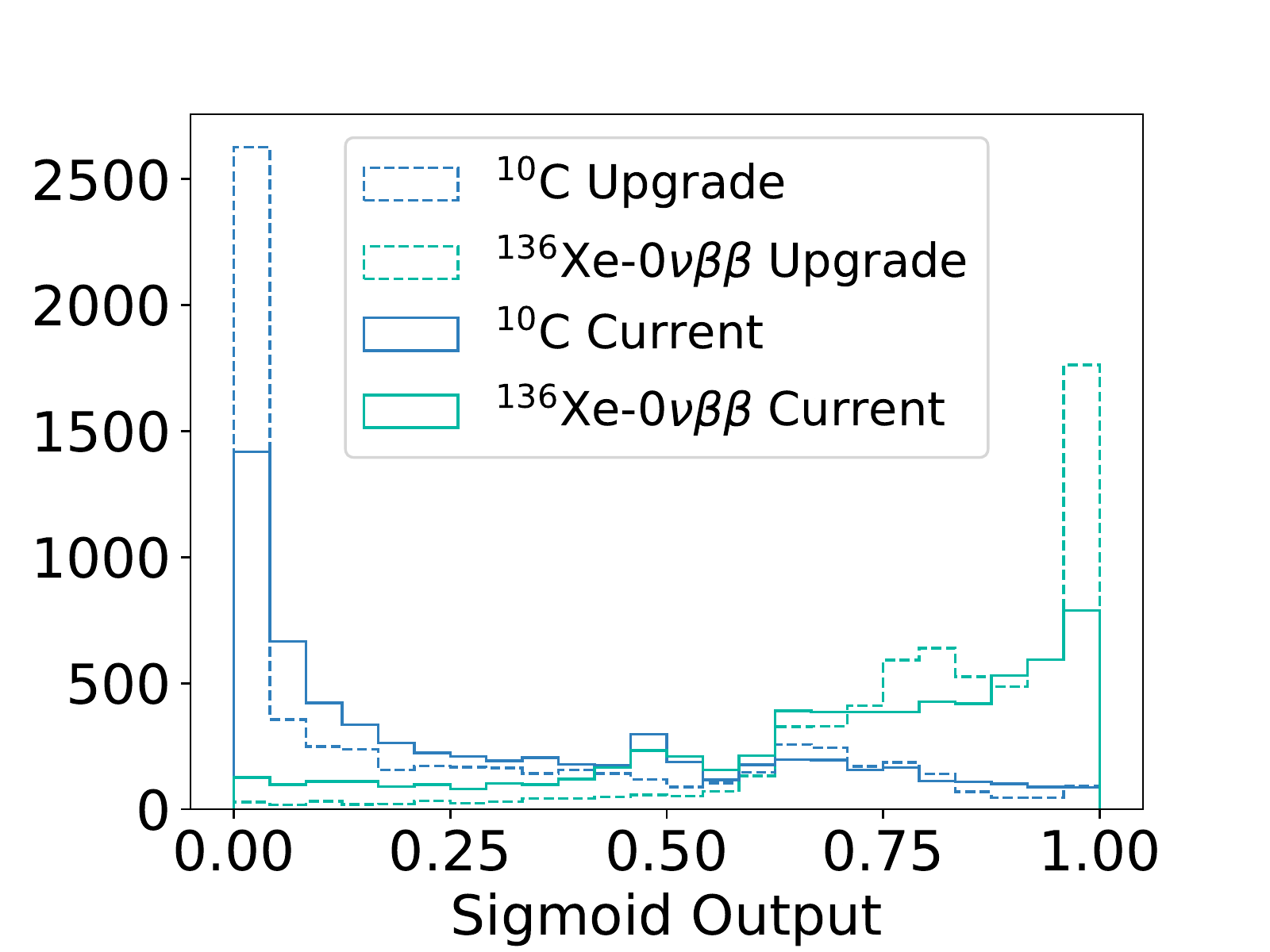}
\caption{Sigmoid output from simulated events isotropically distributed within a 3m-diameter balloon.}
\label{fig:sigmoid}
\end{figure}

The well-trained neural network outputs a single floating point number between 0 and 1 for every input event. This number is the sigmoid output, since it comes out of a sigmoid activation function. The sigmoid output serves as the probability or metric for event classification. If the sigmoid output of a given event is close to 1, it means the event is likely to be a signal event. The value of the sigmoid output used for the classification can be more or less stringent depending on the required signal purity versus signal acceptance. Fig.~\ref{fig:sigmoid} shows the sigmoid output for the current configuration and a possible upgrade scenario.

\begin{figure}[hh!]
\centering
\includegraphics [width=8cm] {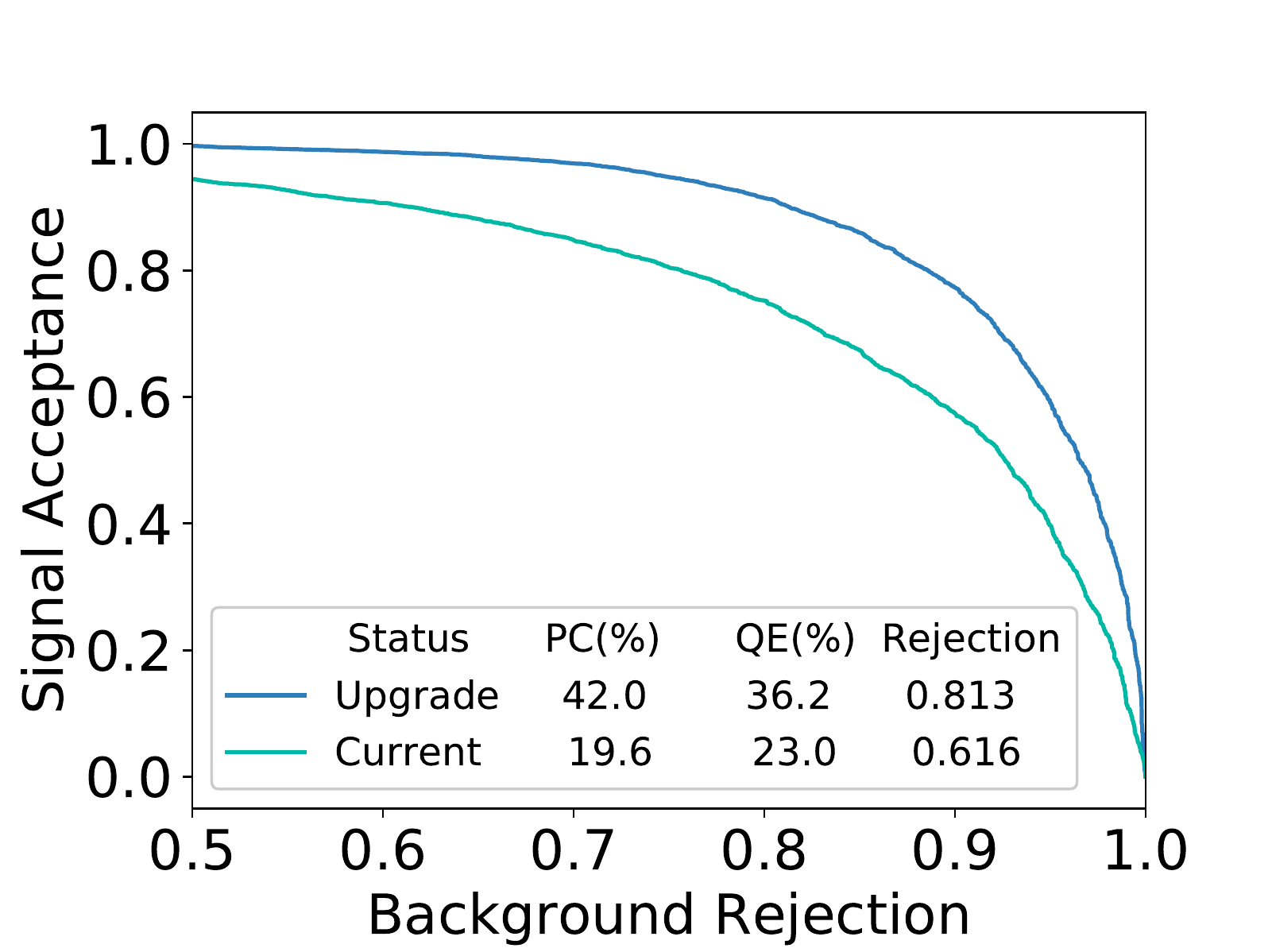}
\caption{ROC curves from simulated events isotropically distributed within a 3m-diameter balloon.  The quoted background rejection assumes 90\% signal acceptance.}
\label{fig:results_curve}
\end{figure}

After the CNN is applied to the validation data set, the value of the sigmoid output cut can be varied to generate a Receiver Operating Characteristic (ROC) curve. For our application, the result is simply the signal acceptance as a function of the background rejection. Fig.~\ref{fig:results_curve} shows the ROC curves for the current configuration and the upgrade scenario. We find that at 90\% signal acceptance we can reject 61.6\% of the \C. For the scenario with increased coverage and QE this increases to 81.3\%. The central events with the standard KamLAND configuration have a 97.7\% rejection at 90\% signal acceptance.

\begin{figure}[hhh!]
\centering
\includegraphics [trim={1.5cm 0cm 2.5cm 2cm},clip, width=1.0\columnwidth] {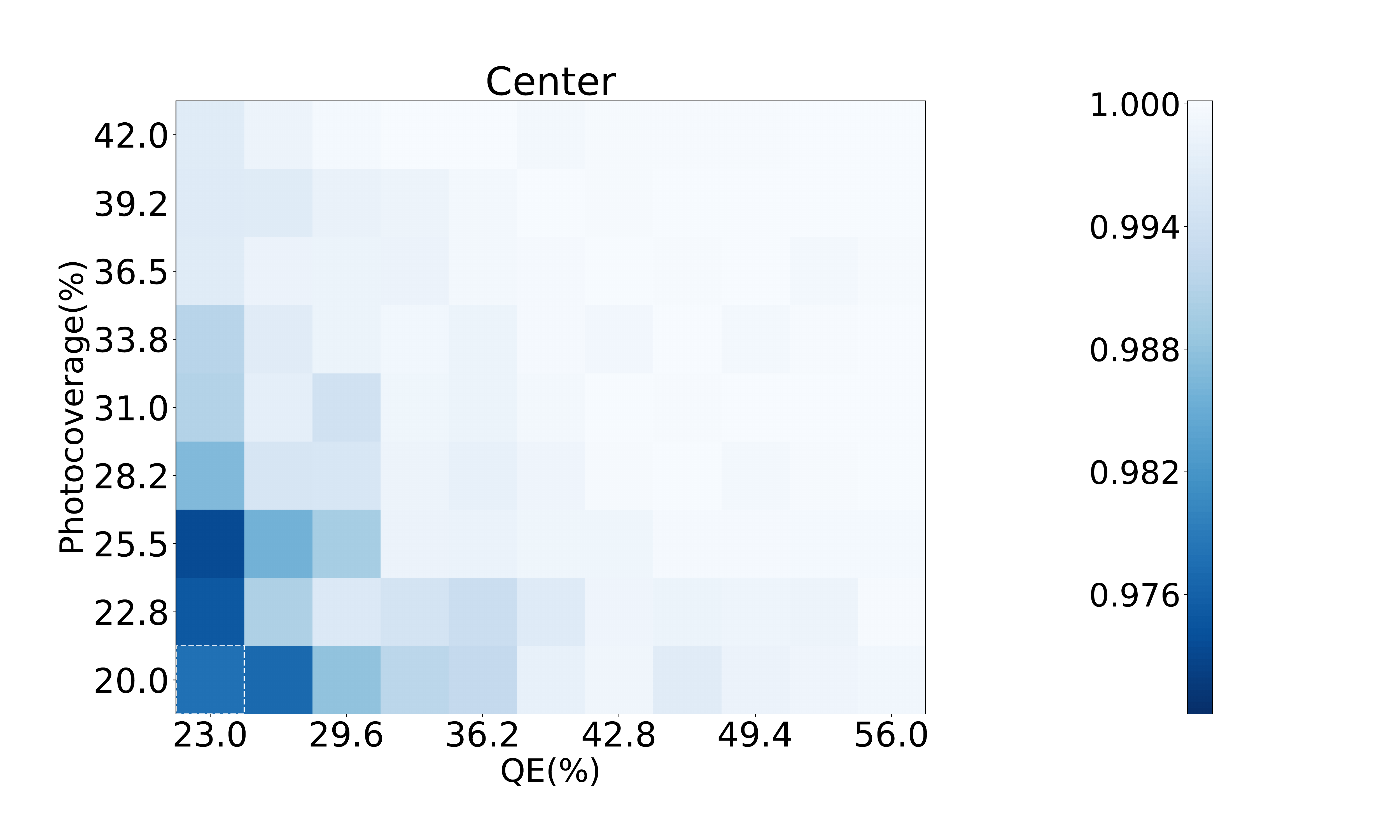}
\includegraphics [trim={1.5cm 0cm 2.5cm 0cm},clip, width=1.0\columnwidth]{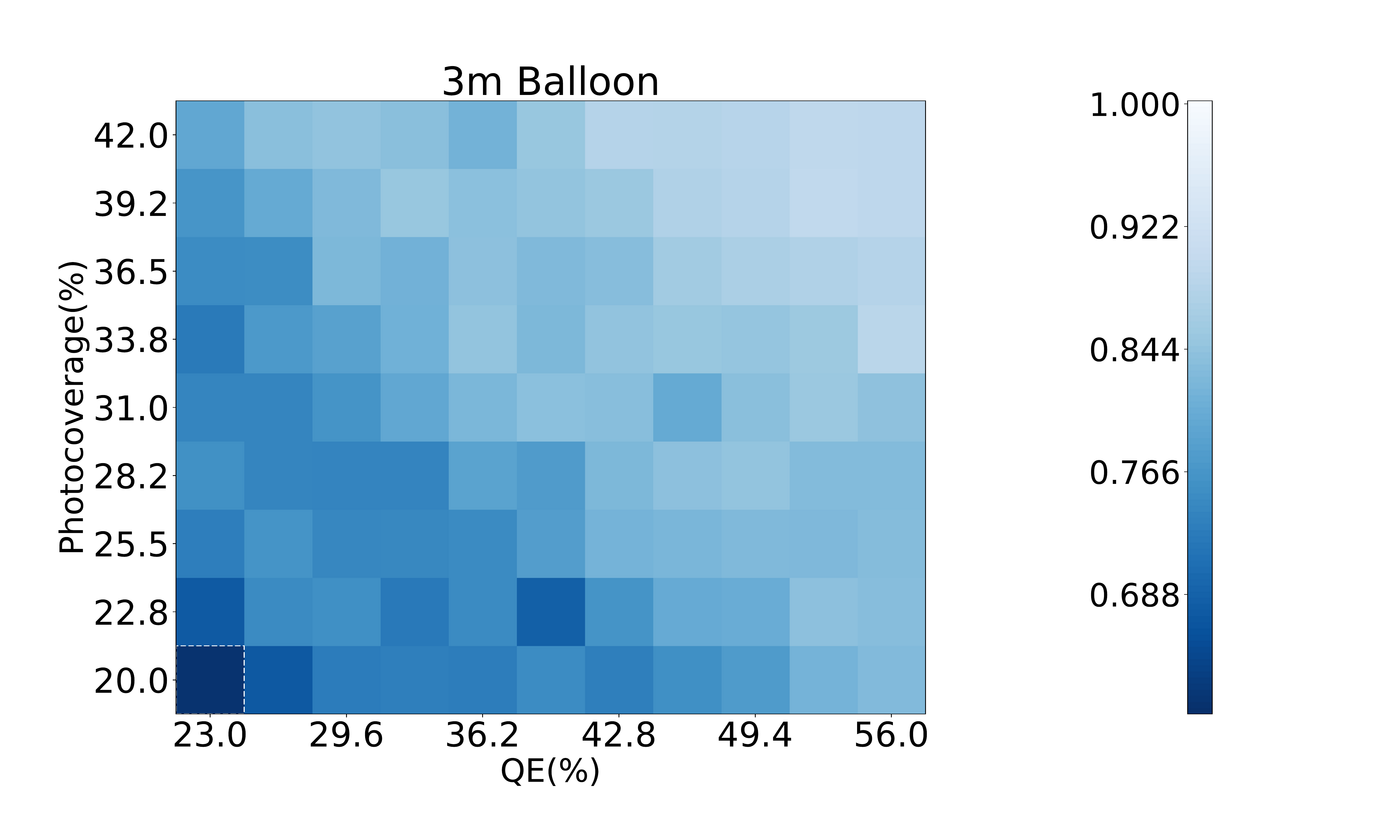}
\caption{Pressure map of the centrally located $^{10}$C events (Top) and the same for $^{10}$C events within a 3m diameter sphere (Bottom). The value in each grid is the Background Rejection Percentage assuming $90\%$ Signal Acceptance. The dashed box indicates the current KamLAND-Zen PMT efficiency.}
\label{fig:results_pressure}
\end{figure}

 Fig.~\ref{fig:results_pressure} shows the pressure maps which scan different QE and photo-coverage configurations for both the central and the 3m sphere events. As expected, the CNN performs best for centrally located events and for higher QE and photo-coverage.  The results also indicate that increasing the total light collected, whether by increasing in QE or photocathode coverage, leads to improved performance.
 
 Within the parameters of this study, we find that it is possible to reach $>$99.98\% discrimination for central events. This indicates that higher isotope concentrations that lead to more centrally distributed \vbb~events are advantageous and this motivates future design studies. We also studied the algorithm with the 3m sphere events and perfect light collection and find 98.2\% rejection at 90\% signal acceptance.
 
 We use the fluctuations observed Fig.~\ref{fig:results_pressure} to understand the uncertainty in the algorithm. This is estimated by calculating at each point in the grid the standard deviation relative to the 4 adjacent neighbors and averaging this value over the pressure map. For central events, the uncertainty is 0.16\%, while the 3m Sphere events give an 1.9\%. 

In order to understand what feature the CNN is using to discriminate, we produced a data set where the Cherenkov light was removed. The results for central events with the standard KamLAND detector configuration indicated a 4\% decrease in the rejection. The 3m sphere events show a 2\% increase in the rejection. This could indicate that Cherenkov light is interfering with the networks interpretation of the rise-time of the scintillation light, however this is also within our estimated uncertainty for the algorithm. As a whole, we find that the Cherenkov signal is not the dominant feature and that the scintillation light topology is driving the event separation. Within the scintillation light topology the photon timing information contains significant discriminating power. Removing spatial information and using a one-dimensional fully connected network with timing as an input we found background rejection drops from 61.6\% to 55.0\% at 90\% signal acceptance.

\section{\label{sec:Conclusion}Conclusion}
Liquid scintillator detectors have been at the heart of many of the great discoveries in neutrino physics and have been a leading technology in the search for \vbb. However, the algorithms used to analyze their data have remained relatively unchanged for decades. In this work, we apply an algorithm from computer vision based on a CNN to extract the fundamentally different physics processes that take place in \C~background events and \vbb~signal events. With a standard detector configuration similar to the current KamLAND detector, we find we can reject 61.6\% of the \C~background with 90\% acceptance of the \vbb~signal. A detector with the same geometry and perfect light collection could achieve 98.2\% rejection. We also find that the performance can be increased to better than 99.98\% for centrally located events. The overall uncertainty of the algorithm is 1.9\%.

These results are a basis for future studies combining machine learning techniques based on CNNs with liquid scintillator detectors. In short order, we intend to move to a spherically symmetric CNN~\cite{s2cnn} and a Bayesian classification that provides a posterior distribution for the classification. In future studies, this algorithm will be applied to other backgrounds with topologies distinct from \vbb~decay.  These include $^{214}$Bi decays on KamLAND 3m balloon and elastic scattering of $^{8}$B solar neutrinos. Solar neutrinos are expected to be the dominant background in SNO+~\cite{snoplus}. We are also exploring algorithms which could move beyond simple classification to particle position and direction reconstruction. These studies are benefiting from an abundance of work being done for other applications both inside and outside of particle and nuclear physics and there are many new avenues to explore.

%KamLAND 3m sphere:61.6\% rejection
%KamLAND center: 97.7\% rejection

%Check on this formatting.
\section*{Acknowledgments}
This material is based upon work supported by the National Science Foundation under Grant Numbers 1554875 and 1806440. We thank Taritree Wongjirad for contributions to earlier incarnations of this project and Jonathan Ouellet for the \C~decay diagram. We also thank Kazuhiro Terao for useful discussions. This work is done in support of the NuDot, THEIA and of course KamLAND experiments and we thank them for their input. The work at the University of Chicago is supported by U. S. Department of Energy, Office of Science, Offices of High Energy Physics and Nuclear Physics under contracts DE-SC0008172 and DE-SC0015367; the National Science Foundation under grant PHY-1066014; and the Physical Sciences Division of the University of Chicago. We thank Mikhail Hushchyn, Eugeny Toropov, and Ilija Vukotic for advice on application of machine learning techniques to image classification. This research was done, in part, using resources provided by the Open Science Grid~\cite{Grid1, Grid2}, which is supported by the National Science Foundation award 1148698, and the U.S. Department of Energy's Office of Science.  This research was also performed with the help of the Shared Computing Cluster at Boston University.

%Check on this formatting.
%\subsection{\label{sec:citeref}Citations and References}
\bibliography{bibliography_ML_2018}{}
%\bibliographystyle{elsarticle-num}
%\bibliography{<your-bib-database>}

\end{document}